\documentclass[twocolumn, showpacs,superscriptaddress,nofootinbib,floatfix, amsfonts, aps]{revtex4}
\usepackage{amsmath}
\usepackage{bm}
\usepackage{graphicx}
\usepackage{mathrsfs}
\usepackage{footmisc}
\usepackage{multirow}
\usepackage{graphicx}
\usepackage{booktabs, ctable}
\usepackage{xcolor, color, framed}

\newcommand{\jpsi}{$J/\psi$ }

\begin{document}
\author{Jiaxing Zhao} 
\affiliation{Department of Physics and Collaborative Innovation Center of Quantum Matter,
Tsinghua University, Beijing 100084, China}

\author{Baoyi Chen}
\affiliation{Department of Physics, Tianjin University, Tianjin 300350, China }

\title{
Strong Diffusion Effect of Charm Quarks on 
$J/\psi$ Production in 
Pb-Pb collisions 
at the LHC}

\date{\today}

\begin{abstract}
We study the $J/\psi$ production based on coalescence model at 
$\sqrt{s_{NN}}=2.76$ and $5.02$ TeV Pb-Pb collisions. With the colliding energy 
increasing from 2.76 TeV to 5.02 TeV, the number of charm pairs is enhanced 
by more than 50\%. However, the ratio of $J/\psi$ inclusive nuclear modification factors 
$R_{AA}^{\rm 5.02 TeV}/R_{AA}^{\rm 2.76 TeV}$ is 
only about $1.1\sim 1.2$. We find that the regeneration of $J/\psi$ is proportional 
to the densities of charm and anti-charm quarks, instead of their total numbers. The charm quark 
density is diluted by the strong expansion of quark gluon plasma, which suppresses the combination 
probability of heavy quarks and $J/\psi$ regeneration. This effect is more important in higher 
colliding energies where QGP expansion is strong. We also propose the ratio 
$N_{J/\psi}/(N_{D})^2$ as a measurement of $c$ and $\bar c$ coalescence probability, which is 
only affected by the heavy quark diffusions in QGP, and does not depend on the inputs such as 
cold nuclear matter effects and cross sections of charm quark production.   
Further more, we give the predictions at the energy of Future Circular Collider 
($\sqrt{s_{NN}}=39$ TeV).  

\end{abstract}

\pacs{25.75.-q, 12.40.Yx, 14.40.Pq, 14.65.Dw}

 \maketitle
A new kind of matter called ``Quark Gluon Plasma" (QGP) is believed to
be produced in the relativistic
heavy ion collisions~\cite{Bazavov:2011nk}. $J/\psi$ has been considered as a probe of this
deconfined matter for more than thirty years~\cite{Matsui:1986dk}.
The color screening and parton inelastic scatterings in
QGP can result in the abnormal suppression of $J/\psi$
production in heavy ion collisions~\cite{Gonin:1996wn, 
BraunMunzinger:2000px, Grandchamp:2002iy,Chen:2017jje}.
The nuclear modification factor $R_{AA}$ is a measurement of the cold and hot medium
effects on charmonium production.
Cold nuclear matter effects include the nuclear absorption~\cite{Gerschel:1988wn}, Cronin effect~\cite{Cronin:1974zm} and
shadowing effect~\cite{Mueller:1985wy,Liu:2016zps}. The first one means that primordially produced charmonium from
parton hard scatterings~\cite{Ma:2017rsu} suffer the
inelastic scatterings with surrounding nucleons before they move out of the nucleus.
This nucleus suppression (``normal suppression") can be neglected at the energies of Large
Hadron Collider (LHC).
Partons may scatter with other nucleons to obtain extra energy
before they fuse into a charm pair (or charmonium). This energy will be inherited by the primordial
$J/\psi$ and shift their transverse momentum distribution. Cronin effect increases with the
number of participants, and can be included by the modification of charmonium initial production
from pp collisions. Parton distributions may also be affected by the surrounding nucleons especially
at the LHC energies. This will change the yields of primordial charmonium and charm quark pairs.

With more and more experimental data published from
Relativistic Heavy Ion Collider (RHIC)~\cite{Adamczyk:2012pw,Adare:2006ns}
and LHC~\cite{ALICE:2013xna,Chatrchyan:2012np}, the nuclear modification factor is enhanced
at higher colliding energies. This
is due to the recombination of $c$ and $\bar c$ quarks in the QGP. At LHC, most of primordially produced
$J/\psi$ are melt by QGP. The recombination of uncorrelated $c$ and $\bar c$ quarks
dominates the $J/\psi$
final yield in nucleus-nucleus collisions, especially at the low $p_T$ region.
For the $p_T$-integrated observables in semi-central and central collisions, one can
safely neglect the primordial production and 
focus on the $J/\psi$ regeneration~\cite{
Zhou:2014kka,Zhao:2011cv,Liu:2009gx, Adam:2015isa}. 

As final $J/\psi$ are mainly from the coalescence of $c$ and $\bar c$ quarks in QGP, $J/\psi$
production is closely connected with the heavy quark evolutions in the expanding
QGP. The elliptic flows of D
mesons at $\sqrt{s_{NN}}=2.76$ TeV is close to the value of light hadrons, which indicate a
kinetic equilibrium of charm quarks before hadronization~\cite{ALICE:2012ab,Song:2011kw}.
Charm quarks expand outside
with the QGP, which reduces its density in phase space.
The effect of charm quark diffusion in
coordinate space becomes more important in the higher colliding energies where QGP expansion is
strong\cite{Zhu:2007ne}.
With higher initial temperature, QGP takes
longer time to cool down where charm quarks will be
distributed in a larger volume~\cite{Liu:2014rsa}.
Therefore, the ratio of $J/\psi$ nuclear modification factors
$R_{AA}^{\rm 5.02TeV}/R_{AA}^{\rm 2.76TeV}$ is only $1.1\sim 1.2$, even the
total number of charm pairs is enhanced by more than 50\% from 2.76 TeV
to 5.02 TeV Pb-Pb collisions.

At the hadronization, partons in the deconfined phase are transformed into hadrons. 
Coalescence model has been
widely used to describe the hadronization process of light
hadrons and heavy quarkonium~\cite{Fries:2008hs,Hwa:2002tu,Molnar:2003ff,Fries:2003vb,Gorenstein:2000ck,Greco:2003vf}. 
Whether $c$ and $\bar c$ quarks 
form into a quarkonium bound state depends on 
both their relative coordinate and momentum, and also the wave functions of charmonium. 
We employ the non-relativistic Schr\"odinger 
equation to obtain charmonium wave functions 
due to the large mass of charm quarks, 
\begin{eqnarray}
\left[{p_1^2 \over 2m_1}+{p_2^2 \over 2m_{2}}+V({\bf r}_1, {\bf r}_2)  \right]\Psi=E\Psi
\label{eq-sch12}
\end{eqnarray}
$m_{1,2}$ and $p_{1,2}$ are the mass and momentum of 
charm (or anti-charm) quark. 
$V({\bf r}_1,{\bf r}_2)$ is the 
heavy quark potential. In the coalescence model, $J/\psi$ is regenerated at 
$T=T_c$ just like light hadrons. 
On the hadronization hypersurface, one can 
neglect the parton color screening effect 
on heavy quark potential~\cite{Asakawa:2003re}, and take $V({\bf r}_1,{\bf r}_2)$ to be the form 
of Cornell potential, $V({\bf r}_1,{\bf r}_2)$=${-\alpha / r}+\sigma r$ 
with $r=|{\bf r}_1-{\bf r}_2|$ to be the relative distance between $c$ and $\bar c$ quarks.  
The charm quark mass $m_1(=m_2)$ and ($\alpha,\sigma$) in Cornell potential are taken as parameters 
here, which can be fixed by fitting the mass of ($J/\psi, \chi_c, \psi\prime$) in vacuum. We obtain 
$m_1=m_2=1.25$ GeV and $(\alpha=\pi/12,\sigma=0.2\ \rm{GeV^2}$). 

The potential in above 2-body Schr\"odinger equation depends on 
the relative distance of two quarks, we can seperate the two-body system into 
a motion of the mass center of two quarks which is just an equation of free motion, and 
their relative motion which is controlled by the potential $V({\bf r})$. 
After introducing a global 
coordinate ${\bf R}=({\bf r}_1+{\bf r}_2)/2$ and relative coordinate ${\bf r}={\bf r}_1-{\bf r}_2$, 
we write the total wave function of two body systems as 
$\Psi({\bf R},{\bf r})=\Theta({\bf R})\psi({\bf r})$.   
Further more, the Cornell potential is an isotropic potential. One can write the equations of 
$\psi({\bf r})$ into a radial part and an angular part, 
$\psi({\bf r})=\varphi(r)Y(\theta,\phi)$. Now the two-body system is simplified as a 
one dimensional problem, 
 \begin{eqnarray}
\left[-{1\over 2m_\mu}({d^2 \over dr^2}+{2\over r}{d \over dr})+V(r) 
+{L(L+1)\over 2m_\mu r^2} \right]\varphi(r)=\varepsilon \varphi(r) \nonumber \\
\label{eq-radialsch}
\end{eqnarray}
where $m_\mu=m_c/2$ is the reduced mass. 
$\varphi(r)$ is the radial wave function. For $J/\psi$, 
the angular momentum quantum number is $L=0$. The radial wave function is 
normalized as 
$\int_0^{\infty} |\varphi(r)|^2 r^2dr = 1$. 

Including the dependence of both relative distance $\bf r$ and relative momentum $\bf p$, 
one can write the Wigner function for 
$c$ and $\bar c$ quarks hadronization 
into a charmonium as~\cite{Zhao:2017gpq,He:2014tga}, 
\begin{equation}
\label{wigner1}
W({\bf r},{\bf p})=\int d^3{\bf y} e^{-i{\bf p}\cdot{\bf y}}\psi\left({\bf r}
+{{\bf y}\over 2}\right)\psi^*\left({\bf r}-{{\bf y}\over 2}\right)
\end{equation}
where $\psi({\bf r})$ is the wave function of $J/\psi$ obtained in Eq.(\ref{eq-radialsch}). 
The Wigner function is normalized as $\int_0^\infty W({\bf r},{\bf p})
d^3{\bf r}{d^3 {\bf p}\over (2\pi)^3}=1$ in the nonrelativistic limit.

Before doing dynamical evolutions of heavy quarks, we give 
the realistic evolutions of QGP produced in Pb-Pb collisions. The QGP turns out 
to be a very strong coupling system, which can be described well 
by hydrodynamic equations~\cite{Kolb:2000sd,Song:2011qa,Zhao:2017yhj}. 
With the assumption of Bjorken expansion for QGP longitudinal expansion, 
we employ the 2+1 dimensional hydrodynamic equations to simulate QGP transverse expansion in 
Pb-Pb collisions at LHC energies, 
\begin{eqnarray}
\partial_\mu T^{\mu v}=0
\label{eq-hd}
\end{eqnarray}
Here $T^{\mu v}$=$(e+p)u^\mu u^v-g^{\mu v}p$ is the energy-momentum tensor, and $e,p,u^\mu$ 
are the energy density, pressure and four velocity of fluid cells, respectively. 
For the equation of state, the deconfined matter is taken as 
an ideal gas of $u$, $d$, $s$ quarks and gluons~\cite{Sollfrank:1996hd}. The hadron phase is 
an ideal gas of all known hadrons and resonances
 with mass up to 2 GeV~\cite{Olive:2016xmw}. There is a first order phase transition between 
two phases. In the mixed phase,  
Maxwell construction is used to obtain the values of variables in Eq.(\ref{eq-hd}). 

From the charge multiplicity, one can obtain the initial 
maximum temperature of QGP at $\sqrt{s_{NN}}=2.76$ TeV 
and 5.02 TeV Pb-Pb collisions to be $T_{0}^{\rm 2.76 TeV}(\tau_0)=485$ MeV 
and $T_{0}^{\rm 5.02 TeV}(\tau_0)=510$ MeV 
in central rapidity bin in the most central collisions (impact parameter b=0)~\cite{Chen:2017duy}. 
$\tau_0$ is the time 
where QGP reaches local equilibrium and starts transverse expansion. 
Its value at RHIC 200 GeV Au-Au collisions and 2.76 TeV Pb-Pb collisions 
are both $\sim 0.6$ fm/c~\cite{Chen:2013wmr}, 
showing weak dependence of the colliding energy $\sqrt{s_{NN}}$. 
Therefore, we also take 
its value to be $\tau_0^{\rm 5.02 TeV}=\tau_0^{\rm 39 TeV}=0.6$ fm/c. 
The initial maximum temperature 
of QGP at $\sqrt{s_{NN}}=39$ TeV Pb-Pb collisions in the Future Circular Collider is extracted as 
$T_0^{\rm 39 TeV}(\tau_0)= 650$ MeV~\cite{Hirano:2010jg,Zhou:2016wbo}.  

$J/\psi$ production from coalescence of $c$ and $\bar c$ quarks is proportional to the densities of 
charm and anti-charm quarks and Wigner function~\cite{Fries:2003kq}, 
\begin{align}
\label{coa}
\frac{dN_{J/\psi}}{d^2{\bf P}_Td\eta} = C\int &{\frac{P^\mu d\sigma_\mu(R)}{(2\pi)^3}
{\frac{d^4 r d^4 p}{(2\pi)^3}}} \nonumber \\
&\times W( r, p)  f_c( {\bf r_1},  {\bf p_1})f_{\bar c}( {\bf r_2}, {\bf p_2})  
\end{align}
where $P^\mu =(P^0,{\bf P})$ is the momentum of $J/\psi$, 
$P^0=\sqrt{m_{J/\psi}^2+{\bf P}^2}$. 
The constant $C$ comes from the intrinsic symmetry with $C=1/12$ for vector 
mesons like $J/\psi$. $f_{c}({\bf r},{\bf p})$ is the charm quark density in 
phase space. The coalescence of charm and anticharm quarks happens on the hadronization 
hypersurface $\sigma_\mu(R)$, where the coordinates $R_\mu$=$(t,{\bf R})$ on the 
hypersurface is constrainted by the hadronization condition: $T(R_\mu)$=$T_c$.

\begin{figure}[htb]
{ $$ \includegraphics[width=0.45\textwidth] {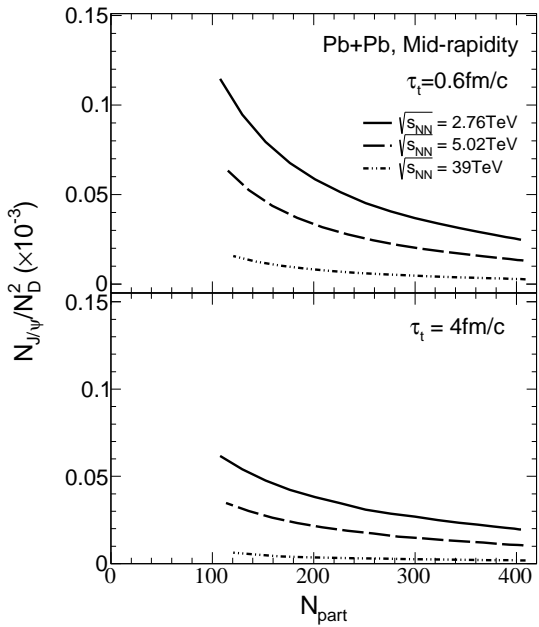}$$
\caption{The ratio of $J/\psi$ yield to the square of D meson yield 
(including all the 
open-charm mesons) as a function of 
number of participants at mid-rapidity in Pb-Pb collisions at different colliding energies. 
The (solid, dashed, dotted) line corresponds to 
the colliding energy $\sqrt{s_{NN}}=$(2.76, 5.02, 39) TeV, respectively. 
Upper panel: Charm quark reach kinetic equilibrium at $\tau=\tau_0$ 
just like light partons. Lower panel: Charm quark is with free streaming at $\tau< 2$ fm/c, 
and reach kinetic equilibrium at $\tau\ge 2$ fm/c.
}
\label{fig-DJPSI}}
\end{figure}

In $\sqrt{s_{NN}}=2.76$ TeV Pb-Pb collisions, the large elliptic flow of D mesons 
supports the kinetic thermalization of charm quarks in QGP~\cite{ALICE:2012ab}. 
However, how and when charm quarks reach kinetic thermalization 
still deserve more quantitative studies~\cite{Beraudo:2015wsd,Beraudo:2014boa,
Beraudo:2010tw,Beraudo:2009pe}. 
Instead of doing realistic evolutions 
of charm quarks which introduce additional 
parameters, for simplicity, 
we assume an instant kinetic thermalization of charm quarks at $\tau=\tau_0$ 
just like light partons. 
The situation of charm quark non-thermalization will 
be discussed in Fig.\ref{fig-DJPSI}, and does not change our main conclusions.
The momentum distribution of charm quarks is    
\begin{eqnarray}
f_{c}({\bf r},{\bf p})=\rho_c({\bf r}) \frac{N^{\rm norm}({\bf r})}
{e^{u^{\mu}({\bf r})p_{\mu}/T({\bf r})}+1}
\end{eqnarray}
where $u^\mu ({\bf r})$ and $T({\bf r})$ are the four velocity and local temperature of 
QGP. $ N^{\rm norm}({\bf r})$ is the normalization factor. $\rho_c({\bf r})$ 
is charm quark spatial density. 
As charm quark mass is very large, it can hardly reach chemical equilibrium in the QGP with a typical 
temperature of $0.2\sim 0.5$ GeV. The total number of charm pairs is conserved. With the assumption of 
charm quark kinetic equilibrium in QGP, its diffusion in coordiate space can be expressed as 
a conservation equation, 
\begin{align}
\partial_\mu (\rho_c u^\mu)=0
\label{eq-charmden}
\end{align}
Obviously, the diffusion of charm quarks in coordinate space depends on the velocity 
$u^\mu$ of QGP. 
For the input of Eq.(\ref{eq-charmden}), as charm pairs are produced mainly from the 
parton hard scatterings, its initial distribution in coordinate space is 
\begin{align}
\rho_c(\tau_0,{\bf x}_T,\eta)=\frac{T_A({\bf x}_T)T_B({\bf x}_T-{\bf b})\cosh\eta}{\tau_0} 
\frac{d\sigma^{c\bar c}_{pp}}{d\eta}
\label{eq-cinit}
\end{align}
where $T_{A(B)}$ is the thickness function of the nucleus A(B). 
${\bf b}$ is the impact parameter, and $d\sigma_{pp}^{c\bar c}/d\eta$ is 
the charm pair differential cross section with rapidity in proton-proton collisions. 
Note that in the colliding energies as LHC, the strong shadowing effect will suppress 
the initial number of charm pairs and change the Eq.(\ref{eq-cinit}). At $\sqrt{s_{NN}}=2.76$ TeV 
and 5.02 TeV, the shadowing effect will reduce around $20\%$ of total charm yields~\cite{Eskola:1998df}. 
This will 
suppress the yield of regenerated $J/\psi$ by $36\%$. However, it does not affect the ratio of 
$J/\psi$ yield over the square of D meson yield in Pb-Pb collisions, $N_{J/\psi}/(N_D)^2$. 
We propose this observable as a mean probability of charm and anti-charm quark 
combination in the expanding QGP. It will be discussed in details later.  

In order to quantitatively calculate the inclusive nuclear modification factor of $J/\psi$ to 
compare with the experimental data, one need the cross sections of $J/\psi$ and charm pairs in 
pp collisions. 
At $\sqrt{s_{NN}}=2.76$ TeV, ALICE Collaboration published the $J/\psi$ inclusive cross section  
to be $d\sigma^{J/\psi}_{pp}/dy=2.3\ \mu b$ at $2.5<|y|<4$ and  
 $4.1\ \mu b$ at $|y|<0.9$~\cite{PereiraDaCosta:2011nb}.  
At $\sqrt{s_{NN}}=5.02$ TeV, we take $d\sigma^{J/\psi}_{pp}/dy=3.65\  \mu b$ 
at $2.5<|y|<4$~\cite{Adam:2016rdg}. 
\begin{figure}[!hbt]
\centering
\includegraphics[width=0.42\textwidth]{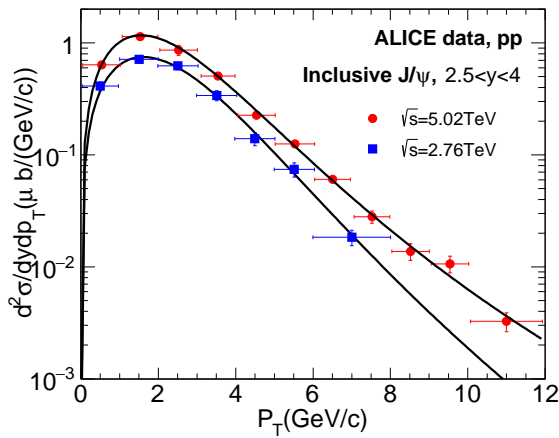}
\caption{
(Color online) The differential cross section of inclusive \jpsi in $\sqrt{s_{NN}}=2.76$ TeV 
(Blue square points) 
and $5.02$ TeV (Red circle points) pp collisions at the forward rapidity. The lines are 
our parameterization with Eq.~\ref{9}. The data are from the ALICE Collaboration~\cite{PereiraDaCosta:2011nb,Adam:2016rdg}.}
\hspace{-0.1mm}
\label{fig-pp}
\end{figure}
The $p_T$-differential cross
sections of inclusive $J/\psi$ in pp collisions at $\sqrt{s_{NN}}=2.76$ TeV 
and $5.02$ TeV are shown in Fig.\ref{fig-pp}. 
It can be parametrized as~\cite{Chen:2016dke},
{\footnotesize
\begin{align}
{d^2\sigma_{pp}^{J/\psi} \over2\pi dyp_Tdp_T}  &={2(n-1)\over2 \pi(n-2)\langle 
p_T^2\rangle^{J/\psi}_{pp}}
[1+{p_T^2 \over (n-2) \langle p_T^2\rangle^{J/\psi}_{pp} }]^{-n} \times
{d\sigma^{J/\psi}_{pp} \over dy}.
\label{9}
\end{align}
}
with n=4.0 and $\langle p_T^2\rangle_{pp}^{J/\psi}$= $7.8\ \rm {(GeV/c)^2}$ at forward 
rapidity for $\sqrt{s_{NN}}$=2.76 TeV. At 5.02 TeV, The parameters are $n=3.9$ and 
$\langle p_T^2\rangle_{pp}^{J/\psi}$= $8.7\ \rm{(GeV/c)^2}$
 at forward rapidity.

For the charm pair production 
cross sections at $\sqrt{s_{NN}}=2.76$ TeV and 5.02 TeV pp collisions, they are taken as 
$d\sigma_{pp}^{c\bar c}/dy(2.76\ \rm TeV)= 0.33$ mb and  
$d\sigma_{pp}^{c\bar c}/dy(5.02\ \rm TeV)= 0.57$ mb at forward rapidities to explain 
the experimental data. Those values are consistent with the inputs of transport 
models~\cite{Chen:2016dke,Chang:2015hqa}. 
We also shift their values upward by 20\% to consider the 
uncertainties of charm production cross sections.

With the information of charm quark evolutions in QGP and also the coalescence probability 
of $c$ and $\bar c$ quarks which is connected with the wave function of produced particles 
(see Eq.\ref{wigner1}), we can calculate the $J/\psi$ production at the temperature $T=T_c$ 
of the phase transition with Eq.\ref{coa}. The experimental data is for the $J/\psi$ inclusive 
nuclear modification factor, which includes the decay contributions from B hadrons. The 
$J/\psi$s from B hadron decays, labeled as ``non-prompt $J/\psi$", contribute around 10\% in the 
total inclusive yields. This fraction $N^{B\rightarrow J/\psi}/N^{\rm inclusive}$ 
almost does not depend on the 
colliding energy~\cite{Zhou:2014kka}. With the prompt and non-prompt $J/\psi$s, one can write the 
inclusive nuclear modification factor as,
 \begin{eqnarray}
R_{AA}={N_{AA}^{c+\bar c\rightarrow J/\psi +g}+N_{AA}^{B\rightarrow J/\psi}\over N_{coll} 
N_{pp}^{\rm inclusive\ J/\psi}}
\end{eqnarray}
where the first and second term in the numerator are 
$J/\psi$ production from coalescence 
of $c$ and $\bar c$ quarks and 
B hadron decay. $N_{coll}$ and $N_{pp}^{\rm inclusive\ J/\psi}$ are the number of binary collisions 
and $J/\psi$ inclusive yield in pp collisions, which can be obtained from the integration of Eq.\ref{9}.

Besides the coalescence of heavy quarks and decay from B hadrons, $J/\psi$ may also come from the 
parton hard scatterings just like charm quarks at the nucleus colliding time $\tau=0$. This is 
called ``primordial production". In the LHC colliding energies, the initial maximum 
temperature of QGP is around $T\approx 3T_c$, which is 
far above the $J/\psi$ maximum survival temperature 
$T_d^{J/\psi}\sim 1.2T_c-2T_c$. The lower and upper limits of $T_d^{J/\psi}$ correspond to the heavy 
quark potential to be $V=F$ (free energy) and $V=U$ (internal energy). In both situations, most of 
the primordially produced $J/\psi$ will be dissociated in QGP. 
In the peripheral collisions, the initial temperature 
of the produced QGP becomes smaller. The primordial 
$J/\psi$ may survive from hot medium and even dominate the final inclusive yields. Therefore, we 
perform our calculations at $N_p\ge 100$ where the initial maximum temperature of QGP 
is $T_0^{\rm QGP}(\rm N_p=100)>2T_c\ge T_d^{J/\psi}$. 

\begin{figure}[htb]
{ $$\includegraphics[width=0.4\textwidth] {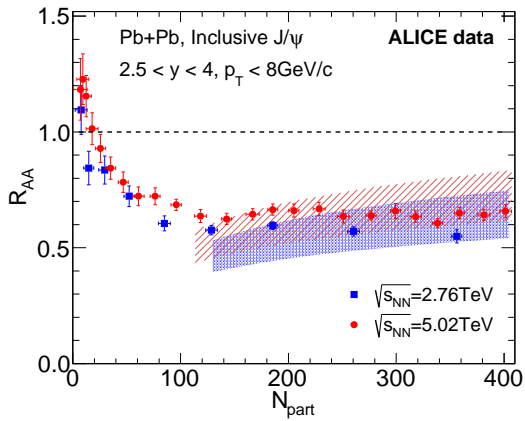}$$
\caption{(Color online) $J/\psi$ inclusive nuclear modification factor $R_{AA}$ as a function of 
the number of participants $N_p$ 
at the forward rapidity in Pb-Pb collisions at $\sqrt{s_{NN}}=2.76$ TeV and 5.02 TeV. 
Lower and upper limits of the bands correspond to the lower and upper limits of 
charm quark cross sections 
in pp collisions. 
Experimental data are from the ALICE Collaboration~\cite{PereiraDaCosta:2011nb,Adam:2016rdg}.}
\label{fig-RAA}}
\end{figure}

\begin{figure}[htb]
{ $$ \includegraphics[width=0.4\textwidth] {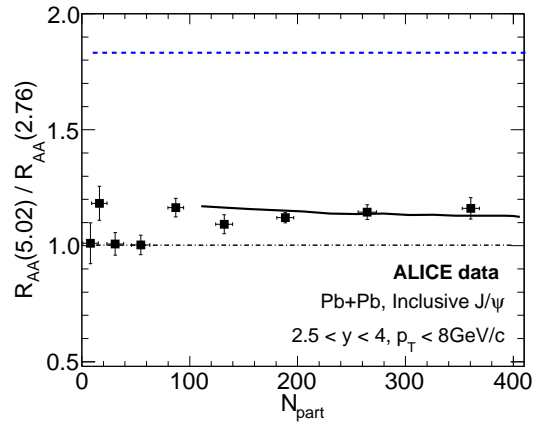}$$
\caption{(Color online) The ratio of $J/\psi$ inclusive nuclear modification factors between 
$\sqrt{s_{NN}}=5.02$ TeV and 2.76 TeV as a function of $N_p$ at the 
forward rapidity in Pb-Pb collisions. 
The black solid line is the calculation with the central values of $J/\psi$ 
and charm pair cross sections. 
The blue dashed line is the ratio of $(\sigma_{pp}^{c\bar c})^2/\sigma_{pp}^{J/\psi}$ 
between 5.02 TeV and $2.76$ TeV. The data are from the ALICE Collaboration~\cite{Adam:2016rdg}.}
\label{fig-DR}}
\end{figure}

The $J/\psi$ inclusive nuclear modification factors at 
2.76 TeV and 5.02 TeV are plotted in Fig.\ref{fig-RAA}. 
From the definition of $R_{AA}$, its value is proportional to the parameters 
of $(\sigma_{pp}^{c\bar c})^2/\sigma_{pp}^{J/\psi}$. 
One can expect a similar enhancement of $J/\psi$ cross section 
from 2.76 TeV to 5.02 TeV just like $\sigma_{pp}^{c\bar c}$, please 
see the cross sections in Section III.
Therefore, the change of the $J/\psi$ and $c\bar c$ production cross sections 
will make $R_{AA}^{\rm 5.02 TeV}/R_{AA}^{\rm 2.76 TeV}\sim 1.8$ depending on the exact values of 
these cross sections (see the blue dashed line in Fig.\ref{fig-DR}). 
It means, 
if we employ the same QGP expansions in Pb-Pb collisions 
at 5.02 TeV as 2.76 TeV for $J/\psi$ regeneration, 
$J/\psi$ $R_{AA}$ will be enhanced by 80\% at 5.02 TeV, 
due to the larger charm and $J/\psi$ cross sections. 
However, with the realistic simulations of QGP expansion at 5.02 TeV, 
$J/\psi$ $R_{AA}$ is only enhanced by around 10\%, 
and the difference of 70\% is due to the stronger 
diffusions of charm qurks in QGP(5.02TeV) 
than in QGP(2.76TeV).
As we point out above, 
the $J/\psi$ production is the integration 
of charm quark density $f_c({\bf r},{\bf p})$, 
which is controlled by Eq.\ref{eq-charmden} and depends 
on the evolutions of QGP. The 
expansion of QGP will affect the dilution of charm quark density. 
Especially, the transverse expansion of QGP is an 
accelerating process. Charm quarks can be ``blown" 
to a larger volume when QGP hadronize, which will 
strongly suppress the charm density at the 
hadronization hypersurface where $J/\psi$ is produced from coalescence process. 
Therefore, the ratio of $R_{AA}^{\rm 5.02 TeV}/R_{AA}^{\rm 2.76 TeV}$ 
is only $1.1\sim 1.2$ in 
semi-central and central collisions, see the 
solid line in Fig.\ref{fig-DR}. Note that in another 
philosophy, where $J/\psi$ is believed to be produced in a temperature 
region $T_c<T<T_d^{J/\psi}$, the strong diffusion of charm quarks in coordinate 
space still suppress the $J/\psi$ regeneration. 

At LHC colliding energies, the final $J/\psi$ are mainly from the coalescence of $c$ and 
$\bar c$ quarks. The large uncertainty of charm quark production cross sections results in 
large uncertainty of theoretical calculations in transport models~\cite{Zhou:2014kka, Zhao:2011cv} and coalescence 
models~\cite{Andronic:2012dm}. It would be better if we can find a observable to describe the effects 
of charm quark diffusion on charmonium production in the expanding QGP. 
In equation Eq.\ref{coa}, we move the cross sections of charm quark production 
to the left hand side. The new observable $N_{J/\psi}/(N_D)^2$ does not depend on 
the cross sections of $J/\psi$ and charm pairs in pp collisions. 
Further, it does not depend on the 
shadowing effect, which can reduce the number of charm quarks. 
Here D meson number  
 equals to charm quark number $N_D=N_c$, independent of 
coalescence or fragmentation for $c\rightarrow D$. Note that $J/\psi$ regeneration 
process reduces only $<1\%$ of total charm numbers and does not affect the 
relation of $N_D=N_c$. 
From the formula 
of $N_{J/\psi}/(N_{D})^2$, it is mainly determined by the $J/\psi$ wavefunction and 
the evolutions of charm quarks which includes the information of bulk medium expansions. 
The combination probability 
of \emph{one} $c$ and $\bar c$ quark becomes smaller if QGP expansion is stronger. 
The value of $ N_{J/\psi}/(N_D)^2$ 
decreases with $N_p$ and the colliding energy $\sqrt{s_{NN}}$ in Pb-Pb collisions, see the lines 
in Fig.\ref{fig-DJPSI}. 

In the above calculations, we assume that charm quarks reach kinetic equilibrium at 
$\tau_0$ for simplicity when QGP starts transverse expansion. However, from realistic 
dynamical evolutions of heavy quarks in the heavy ion collisions~\cite{He:2011yi}, 
charm quarks may be 
thermalized in a few fm/c at the LHC energies. We assume a free motion for 
charm quarks before the time scale $\tau_t=4$ fm/c. 
After this time scale, charm quark momentum reaches kinetic equilibrium 
and their motion is controlled by Eq.\ref{eq-charmden}. Non-thermalization effect does 
not change our conclusions that $J/\psi$ regeneration is suppressed by the spatial diffusion of 
charm quarks in the expanding QGP, see the lower panel of Fig.\ref{fig-DJPSI}. 

\begin{figure}[htb]
{ $$\includegraphics[width=0.4\textwidth] {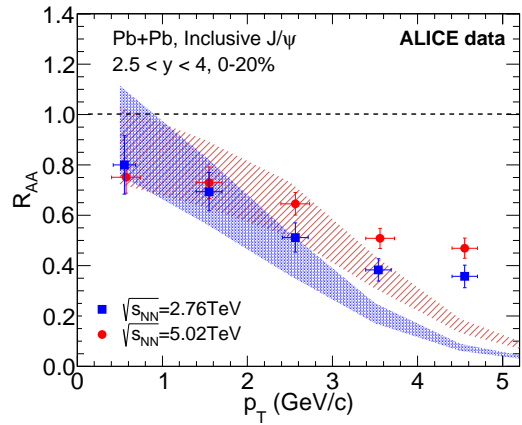}$$
\caption{(Color online) The $J/\psi$ inclusive nuclear modification factor $R_{AA}$ as a function 
of transverse momentum $p_T$ at $\sqrt{s_{NN}}=$2.76 TeV and 5.02 TeV at forward
 rapidity and 0-20\% centrality in Pb-Pb collisions. 
Theoretical bands correspond to the uncertainties of charm quark cross sections at 2.76 TeV and 5.02 TeV. 
Experimental data are from the ALICE Collaboration~\cite{PereiraDaCosta:2011nb,Adam:2016rdg}.}
\label{fig-RAApt}}
\end{figure}

The strong diffusion of charm quarks 
also affects the shape of the transverse momentum distribution of $J/\psi$ $R_{AA}$.  
In Fig.\ref{fig-RAApt}, we calculate the $J/\psi$ inclusive $R_{AA}(p_T)$ in the centrality 0-20\% 
in forward rapidity. Absent of the primordial production which 
dominate the inclusive yield in high $p_T$ region, we underestimate the $R_{AA}$ at $p_T>4$ GeV/c. 
In the region of $p_T<4$ GeV/c, both experimental data and theoretical calculations show a 
``shift" behavior of $R_{AA}$ toward larger $p_T$ region from 2.76 TeV to 5.02 TeV. 
In $\sqrt{s_{NN}}=5.02$ TeV, the 
expansion of QGP is stronger, which pushes charm quarks to a larger transverse momentum region. 
$R_{AA}^{\rm 5.02 TeV}$ is almost the same with $R_{AA}^{\rm 2.76 TeV}$ at $p_T\approx 1$ GeV/c, 
but $R_{AA}^{\rm 5.02 TeV}$ is enhanced at $p_T\approx 3$ GeV/c due to the larger velocity of 
fluid cells in 5.02 TeV. 

In summary, 
we employ the coalescence model to study $J/\psi$ production from the combination of charm quarks 
at the hadronization of QGP in Pb-Pb collisions at $\sqrt{s_{NN}}=2.76$ TeV and 5.02 TeV. From the 
comparison of charm quark cross sections and $J/\psi$ inclusive $R_{AA}$ at these two colliding energies, 
we find that even the number of charm quarks is enhanced by more than 50\% from 2.76 TeV to 5.02 TeV, 
the ratio $R_{AA}^{\rm 5.02 TeV}/R_{AA}^{\rm 2.76 TeV}$ is only $1.1\sim 1.2$. 
$J/\psi$ production is connected with the 
number of charm pairs and also their diffusions in the expanding QGP. 
In higher colliding energies such as 
$\sqrt{s_{NN}}=$5.02 TeV and 39 TeV, the initial energy density of QGP 
becomes larger, and QGP expansion lasts longer compared with the situation of 2.76 TeV. 
Strong diffusion of charm quarks in QGP reduces the probability of $c$ and $\bar c$ quark coalescence 
at the hadronization hypersurface ($T=T_c$). This effect also shifts 
the produced $J/\psi$ to a larger $p_T$ 
region, see $R_{AA}(p_T)$ in Fig.\ref{fig-RAApt}. 
Further, We propose an observable of $N_{J/\psi}/(N_D)^2$ as a measurement 
of charm quark coalescence probability in QGP. It does not depend on the charm quark cross sections and 
cold nuclear matter effects (such as shadowing effect). It is dominated by the evolution history of 
charm quarks in QGP and also the wave function of the produced particle ($J/\psi$), which makes it 
a clean probe to study the charm quark evolutions 
and $J/\psi$ regeneration in heavy 
ion collisions.

\vspace{0.2cm}
\appendix {\bf Acknowledgement}: B. Chen is supported by NSFC under Grant No. 11547043. 
J. Zhao is supported by NSFC and MOST under Grant Nos. 11335005, 11575093, 2013CB922000 and
2014CB845400.

\end{document}